\documentclass{birkart04}
 \theoremstyle{definition}
 
 \theoremstyle{remark}

 \numberwithin{equation}{section}

\begin{document}
%
%
%
%

\title[Heteroepitaxial growth]
 {Off-lattice Kinetic Monte Carlo simulations of strained heteroepitaxial
  growth}
\author[Biehl]{Michael Biehl}
\address{%
University Groningen, Institute for Mathematics and Computing Science  \\
P.O. Box 800, 9700 AV Groningen, The Netherlands}
\email{biehl@cs.rug.nl}

\author[Much]{Florian Much} 
\address{Universit\"at W\"urzburg, Institut f\"ur Theoretische Physik \\
 Am Hubland, D-97074 W\"urzburg, Germany}
\email{much@physik.uni-wuerzburg.de}

\author[Vey]{Christian Vey}
\address{Universit\"at W\"urzburg, Institut f\"ur Theoretische Physik \\
 Am Hubland, D-97074 W\"urzburg, Germany}
\email{vey@physik.uni-wuerzburg.de}




\begin{abstract}
   An off-lattice, continuous space Kinetic Monte Carlo (KMC)
   algorithm is discussed and applied in the investigation  
   of strained heteroepitaxial crystal growth.
   As a starting point, we study a simplifying $(1+1)$-dimensional  
   situation with  inter-atomic interactions given
   by  simple pair-potentials. 
   The model exhibits  the 
   appearance of strain-induced misfit dislocations at 
   a characteristic film thickness. In our KMC simulations
   we observe a power law dependence of this critical thickness 
   on the lattice misfit, which is in agreement
   with experimental results for semiconductor
   compounds.
   We furthermore investigate the emergence of strain induced 
   multilayer islands or {\sl Dots\/} upon an adsorbate wetting layer
   in the so-called Stranski-Krastanow (SK) growth mode.
   At a characteristic kinetic film thickness, a transition from monolayer 
   to multilayer island growth occurs.  
   We discuss the microscopic causes of the SK-transition  and its
   dependence on the model parameters, i.e. lattice misfit, growth rate,
   and substrate temperature.
\end{abstract}

\maketitle

\section{Introduction} \label{introduction}

Atomistic 
models of crystal growth in Molecular Beam Epitaxy and similar techniques
continue to attract considerable interest, see for example \cite{villain,krug,newman,
kotrla} for overviews, a very brief introduction is 
given in  \cite{biehl}, this volume.
Many interesting aspects  of epitaxial growth 
can be discussed for the case of homoepitaxy where only one material is 
present.
This includes the deposition of, for instance, a metal or a
semiconductor compound on a matching substrate crystal. 

The term heteroepitaxial crystal growth refers to situations, where
the deposited adsorbate material differs from the substrate.  
The mismatch can be quite fundamental, 
as for instance in the growth of organic films upon metal substrates.
Perhaps the most frequent and, certainly, the conceptually simplest
situation is that where adsorbate and substrate
would crystallize in the same type of lattice,
but with slightly different lattice spacings. 

The latter case is of particular relevance in the fabrication
of semiconductor heterostructures for potential
technical applications, such as modern 
opto-electronic or storage devices. Prominent examples
are the deposition of Ge on Si, InAs on GaAs, or CdTe on ZnSe.
Frequently, the aim is to produce a smooth adsorbate film
of a well defined thickness on a given substrate.  
However, the misfit induced strain can perturb the lattice 
structure and result in, e.g., dislocations which
reduce the quality of the film drastically.  

On the other hand, mechanisms of strain relaxation can also be
exploited in epitaxial growth. The most prominent example is perhaps
the self-organized formation of multilayered islands
in the so--called  Stranski-Krastanov growth mode \cite{villain}.
The ideal process would yield dislocation-free 
three-dimensional objects of well defined size and shape 
which furthermore display spatial ordering. 
These so--called Quantum Dots can capture single or few electrons
and may play the role of {\sl artificial atoms\/} in a
novel type of solid-state-lasers, for example. For a recent overview
of the many interesting experimental and theoretical aspects of
Quantum Dot physics, see \cite{QD}. 

Besides the obvious technological relevance of heteroepitaxial
growth, it is also highly interesting from a theoretical point
of view.  The self-assembly of Dots might be considered a prototype
example of self-organized ordering processes.
Heteroepitaxy remains furthermore a challenge in the design of growth 
models and the development of novel simulation techniques. 

The natural tool for mismatched heteroepitaxy appears to
be Molecular Dynamics (MD) \cite{md} or quantum mechanical variations thereof,
e.g.\  \cite{parrinello}.
This very attractive but computationally demanding
approach is discussed in the general context
of epitaxial growth in \cite{albe} (K. Albe, this volume),
see \cite{dong} for an example application in the context
of dislocation formation. 
Whereas modern computers allow for the simulation of
very large systems, the most serious limitation of the MD 
technique remains: The physical time intervals that can be targeted are
generally quite small, e.g. on the order of picoseconds.
MBE relevant time scales of seconds or even minutes
do not seem feasible currently, even when applying 
highly sophisticated acceleration techniques as in \cite{voter}.

In  simplifying lattice gas models, the particles
can only be placed exactly at  pre-defined sites.
Nevertheless, it is possible to incorporate certain misfit effects
into such models.  In fact, some of the earliest KMC simulations
were performed in the context of strained heteroepitaxial crystal
growth \cite{madhukar}.  
So--called {\sl ball and spring\/}
models consider  additional harmonic interactions,
i.e.\ {\sl springs\/}, between adsorbate or substrate
atoms at neighboring lattice sites, see \cite{madhukar}
and \cite{khor,lam} for more recent examples.
In other approaches, continuum elasticity
theory is applied to evaluate the contribution and influence 
of strain for a given configuration of the lattice \cite{meixner} 
(and references therein).

However, traditional lattice based models cannot cope with the 
emergence of defects or dislocations in a straightforward fashion.
In the following we will discuss an off-lattice model and
the corresponding Kinetic Monte Carlo (KMC) technique,
which allows for continuous particle positions and in which
strain results directly from the atomic interactions.
Schindler and Wolf have essentially formulated the model and
simulation method in \cite{schindler,schindlerwolf}, and
some of the results presented here have been published
previously in \cite{epldislo,eplstranski,nato}. For a detailed
description of the concept and its applications  see also \cite{muchdiss}.

The method incorporates the essential features of hetero-epitaxy in a natural way, 
and, at the same time, enables us to perform simulations over
reasonable physical time scales.  
In contrast to earlier realizations of similar concepts, see e.g.\
\cite{schindler,schindlerwolf,kew,faux}, we are also able to 
simulate the growth of rather thick films over a wide range of
misfits. 

In section \ref{model} we will define the model
and describe the simulation technique. We briefly 
discuss the role of misfit dislocations as a mechanism
for strain relaxation and present our simulation results 
in section \ref{dislocations}. Section \ref{stranski} 
summarizes our investigations of the
Stranski-Krastanov growth mode and in section \ref{outlook}
we conclude with a brief  outlook on perspective investigations.  

\section{Model and method}  \label{model}
In the following, we address quite fundamental and general 
aspects of heteroepitaxy, rather than the 
properties of specific material systems. 
Hence, the primary goal is to develop a fairly simple model
which still captures the essential features of a 
mismatched growth situation. 

Following several previous studies
\cite{schindler,schindlerwolf,kew,faux}, we
choose a pair potential ansatz to model interactions 
between the particles. To begin with, we consider
a Lennard-Jones (LJ) system \cite{chemistry}.
A strong repulsive term at small distances competes with an
attractive contribution in the potential 
\begin{equation}
           \label{lj} \textstyle
                   U_{ij} (U_o,\sigma) \, = \, 4 \, U_o \, \left[ \,
                  \left( \frac{\sigma}{r_{ij}} \right)^{12}  \, - \,
                  \left( \frac{\sigma}{r_{ij}} \right)^{6}  \, \right].
\end{equation}
The relative distance $r_{ij}$ of particles $i$ and $j$ varies continuously
with their position in space.
By choice of the parameters $U_o$ and $\sigma$ we can specify the different
material  properties in our model: interactions between two substrate or
adsorbate atoms are given by the 
sets $\left\{U_s, \sigma_s\right\}$ and $\left\{U_a, \sigma_a\right\}$,
respectively.  In principle, an independent set of parameters would define the
interaction of substrate with adsorbate atoms. In order to keep the
number of parameters small, we follow a standard approach and  set
$U_{as} = \sqrt{U_s U_a}, \sigma_{as} = \left(\sigma_s + \sigma_a\right)/2$ for
the inter-species potential.

The potential energy of two isolated atoms that interact via (\ref{lj}) 
is minimal for  $r_{ij} = 2^{1/6} \sigma$. Accordingly, the lattice 
spacing in a Lennard-Jones crystal is proportional (and close) to
$\sigma$ as well. The relative lattice misfit in our model is therefore
directly controlled by choice of $\sigma_s$ and $\sigma_a$:
\begin{equation} \label{misfit}
 \epsilon \, =  \left(\sigma_a - \sigma_s\right)/{\sigma_s}.
\end{equation}
Essential features and several principled questions can already be 
studied in the simplifying framework of  $(1+1)$-dimensional growth, where
particles are deposited on a one-dimensional surface. 
For the simulations presented here we 
have prepared, at least, 6  layers of substrate particles, each containing $L$ 
particles.  The system sizes vary between $L=200$ and $L=800$  in the following, 
depending on the computational costs of the considered problem.
We assume periodic boundary conditions in the lateral direction
and fix the particle positions in the bottom layer in order to
stabilize the crystal.

In our simulations of the growth kinetics, adsorbate particles are 
deposited at a constant rate $R_d$  as measured in monolayers per second $(ML/s)$. 
Desorption events will be completely suppressed here as they
would be very unlikely anyway.  
Only adsorbate particles at the surface are considered 
mobile in the sense
that they perform diffusion hops and the like.  
However, bulk crystal adsorbate
and substrate atoms will be due to relaxation processes which modify their
spatial positions and relative distances without changing the topology. 

The most distinctive feature of the method is the treatment
of thermally activated processes, for which we assume Arrhenius like 
rates, see \cite{villain,krug,newman,biehl}.
 In lattice gas models, particles are moved from one site to another
with a rate which usually depends on a small neighborhood only.
In the off-lattice approach
any event concerns the entire configuration and its rate depends
on  all particle coordinates, in principle.  
At a given time, the  system resides in a local minimum of
the total potential energy  of the $n$ particles:
\begin{equation} \label{Etot}
 E_{tot} \, = \, \sum_{i=1}^n \,\,\, \sum_{j=i+1}^n \, U_{ij}.
\end{equation} 
A possible event $k$ takes the system to one of the neighboring minima 
in configuration space  and for any such change the corresponding
energy barrier $E_k$ and rate $R_k$ has to be evaluated
by means of a so-called Molecular Statics calculation, see
\cite{schroederwolf,barkema1} for examples and further references.

    \begin{figure}[t]
    \begin{center}
    \fbox{ \includegraphics[scale=0.40]{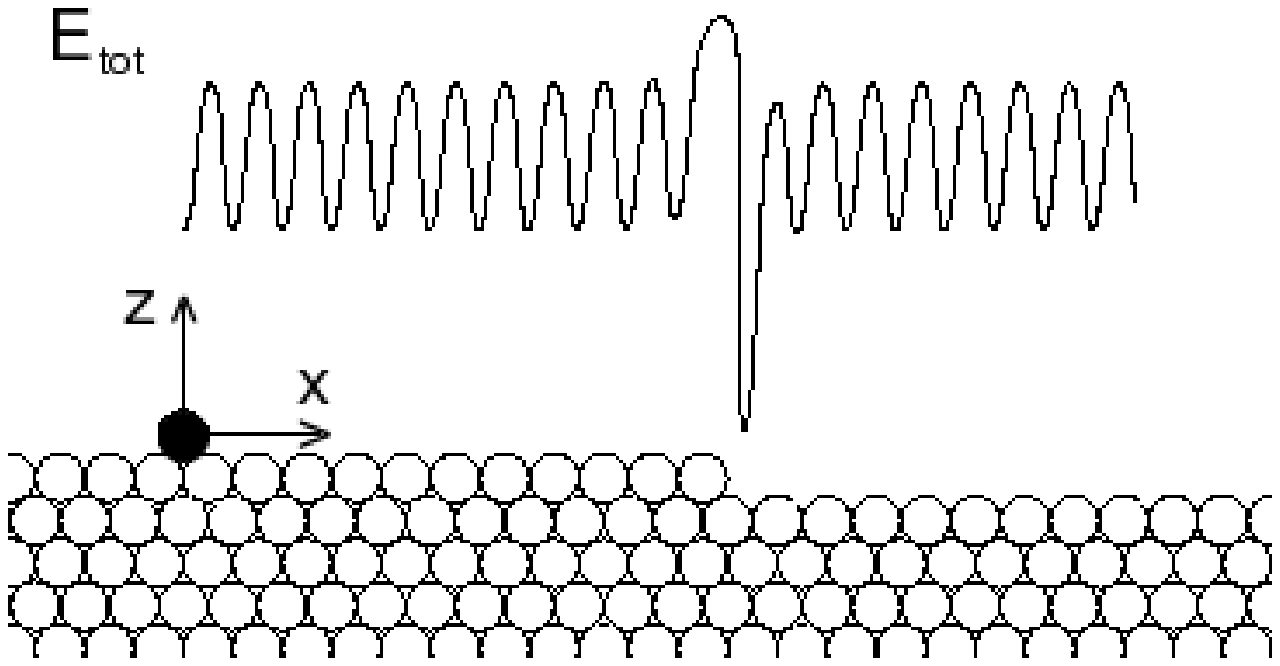}}
     \caption{\label{schwoebel}
      Illustration of the Molecular Statics calculations in 
      $(1+1)$ dimensions. As the (black) adatom is
      virtually moved to the right, along the $x$-direction, the total
      energy of the system is locally minimized with respect to all other
      particle coordinates,
      including $z$ for the moving adatom. The resulting potential energy surface
      displays an essentially periodic shape along the terraces and a
      pronounced Schwoebel barrier for moves across the edge.}
    \end{center}
    \end{figure}

In a most general setting the identification of potential events, i.e.
the relevant energy minima, can be difficult
\cite{barkema1,barkema2,malek}.
The $(1+1)$-dimensional situation
simplifies these calculations to a large extent.

Figure \ref{schwoebel} illustrates the procedure: The single
particle on top of the terraced surface is virtually
moved to the right, along the $x$-coordinate in the illustration.
For every fixed value of $x$, the total energy (\ref{Etot}) 
can be (locally) minimized with respect to the moving atom's $z$-coordinate and the 
positions of all other atoms including the substrate. In doing so, one
can evaluate a  potential energy surface (PES) 
as displayed in Fig.\ \ref{schwoebel}.
Of course it is not necessary to scan the PES continuously for the Arrhenius
rates, as  only the energies in the initial state and the transition state  
are relevant.
The latter corresponds to a maximum in the figure,
i.e. a saddle point in the energy as a function of  all coordinates.
In more complex situations like $(2+1)$-dim. growth 
it is also possible to determine saddle points in a PES by iterative gradient based methods
\cite{muchdiss,barkema1,barkema2,malek}.  
However, the 
identification of the physical path from one minimum to the next
can be much more involved in general, e.g.\ for concerted moves
in $(2+1)$ dimensions.

Note that the $(1+1)$-dim.\  system displays a strong
Ehrlich-Schwoebel effect \cite{villain,krug}: hops from a terrace
are hindered by a  barrier which relates to 
the very weakly bound transition state right at the edge. 
The effect is also present, but
much weaker in a $(2+1)$-dim. fcc-system, for instance. There, 
a descending particle can follow a more favorable path {\sl between}
neighboring particles at the edge. 

For large misfits, exchange processes at terrace or island 
edges may become more frequent than downward hopping diffusion.
We have checked that
for the values of $\epsilon$ considered in the following, exchange diffusion
can be neglected, in general. The precise barriers, however, depend
in a subtle way on the misfit, the island size, and the properties
of the actual interaction potential. 
For details we refer to  \cite{muchdiss} and forthcoming publications. 
    \begin{figure}[t]
     \fbox{ \includegraphics[scale=0.33]{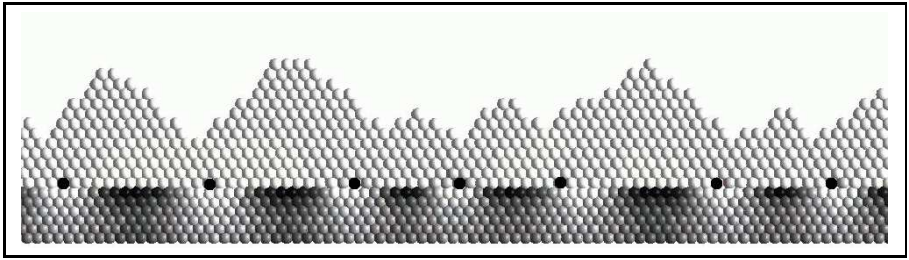} }
     \caption{ \label{manydisl}
     Dislocations: a film grown in the model without downhill
     funneling \cite{epldislo} for a large misfit
     of $\epsilon=+10\%$, section of a larger system with six layers of
     substrate. 
     The lighter a particle is displayed, the smaller is
     its average distance to the nearest neighbors. 
     Dislocations form right at the interface of
     adsorbate and substrate, as marked by the additional
     filled circles. Note that the elastic interaction
     with the film also affects the substrate.  }
    \end{figure}

Further simplifications may be used here. 
First, because the LJ-potential (\ref{lj})
 is very weak for distances $r_{ij} > 3 \sigma$, one can
effectively cut-off the interactions and consider only particles
in a corresponding neighborhood region when evaluating the PES
of a particular event. 
The so-called {\sl frozen crystal approximation\/} 
should be employed with care:
in this scheme all coordinates but those of the moving atom are
considered fixed, which significantly simplifies the saddle point
search.  
Although the scheme appears to impose a drastic restriction,  
it has been reported by various 
authors  that its main effect in the LJ-system 
is a uniform shift  of all barriers by  roughly 10\%
\cite{schroederwolf,schindler,schindlerwolf,muchdiss}.

In our simulations, exchange diffusion and multiple
jumps have been neglected. We consider only hops to
the left or right neighboring minimum in the PES. 
Given a configuration of the system, we set up a catalogue of
all these diffusion events and evaluate their rates 
\begin{equation} \label{arrhenius} 
  R_k \, = \nu_o \, \exp\left[- \frac{E_k}{k_B T}\right] 
 \qquad \mbox{~with~} \qquad \nu_o = 10^{12} /s
\end{equation}
with $E_k$ obtained as described above. 

For simplicity, we choose a constant pre-factor
or {\sl attempt frequency} $\nu_o$ for all diffusion processes,
see \cite{biehl} for a brief discussion and references.
Next, one of the diffusion or deposition processes
is chosen with a probability proportional
to its rate.  In order to facilitate an efficient book-keeping
and fast search,  events are 
stored in a binary search tree, see \cite{newman} and references therein.   
After performing the particular process, the catalogue of rates
has to be updated
and re-evaluated. Obviously this task consumes a large portion
of the computing power. Nevertheless, the rejection-free  method
is advantageous over a rejection-based simulation \cite{biehl}.

Approximations and numerical inaccuracies, and even more so 
the use of the {\sl frozen crystal\/} picture, result 
in deformations of the crystal and the accumulation of
artificial strain energy.  In order to avoid this effect, a relaxation of
the system should be performed after each event, in the sense that the
configuration is taken to the nearest local minimum of $E_{tot}$.
In practice we perform after each diffusion or deposition
a {\sl local\/} relaxation within a radius $3\sigma$ about the  
location of the event.  The full {\sl global\/} procedure is applied 
only after a certain number of events, which is 
decreased whenever the relaxation changes the barriers  by more than, say,
a few percent.    

In the following sections we discuss applications in the context
of two different mechanisms of strain relaxation. Their peculiarities
require certain extensions and modifications. However, the basic concepts
and essential ingredients have been outlined above.

\section{Formation of dislocations}  \label{dislocations}

If the misfit is relatively small in heteroepitaxial growth,
one frequently observes the initial growth of
a {\sl pseudomorphic\/} film. Here the substrate
determines the lateral lattice constant also in the adsorbate
and strain energy accumulates in the growing film.
On the other hand,
for very thick adsorbate films far away from the substrate, 
one clearly expects to find a lattice structure and
spacing as in the undisturbed bulk material. 
The ultimate relaxation of strain is through misfit
dislocations, lattice defects which allow
the adsorbate to  assume its natural lattice structure,
eventually. 
Here we address the question of how the 
transition from strained pseudomorphic to relaxed adsorbate growth
occurs and how the lattice spacing evolves with the film thickness.

In the frame of our model we choose the same potential depth for
all types of interactions, which defines the energy scale. 
By identifying  $U_a=U_s=U_{as}$ with the value  $U_o=1.3125 eV$,
for instance, we obtain a barrier of $0.90 eV$ for surface diffusion
in the case of zero misfit.
The substrate temperature was set to 
$T = 0.03 U_o / k_B \approx 460K$
which is well below the melting
temperature of a monoatomic Lennard-Jones crystal. 
All results presented in this section were obtained
with the deposition rate  $R_d = 1 ML/s$.

Dislocations can be identified by constructing 
Voronoy polyhedra and searching for particles with a coordination
number different from six in the bulk. Further information is
obtained from a Burgers construction, see \cite{epldislo,muchdiss} for
details.

In our earlier investigations presented in \cite{epldislo}
we considered positive and negative values of the misfit parameter in the 
range $-15\% \leq \epsilon \leq + 11\% $.
For large absolute values of $\epsilon$ 
the system displays the formation of misfit dislocations
right at or very close to the substrate/adsorbate interface, cf.
Fig. \ref{manydisl}.
The slightly oversimplifying picture is that, initially, 
separated islands or mounds grow on the substrate and dislocations 
emerge where they meet. These are then overgrown by the 
deposited material.  Diffusion of dislocations by concerted
moves of the surrounding particles are highly improbable
in the LJ-system and for the low temperatures considered here.
We find a mean distance between dislocations  very close to $1/\epsilon$   
which directly reflects the \textit{relative periodicity}
of the two lattices involved. 

\begin{figure}[t]
    \begin{center}
     \fbox{ 
     \makebox[1mm]{}
     \includegraphics[scale=0.30]{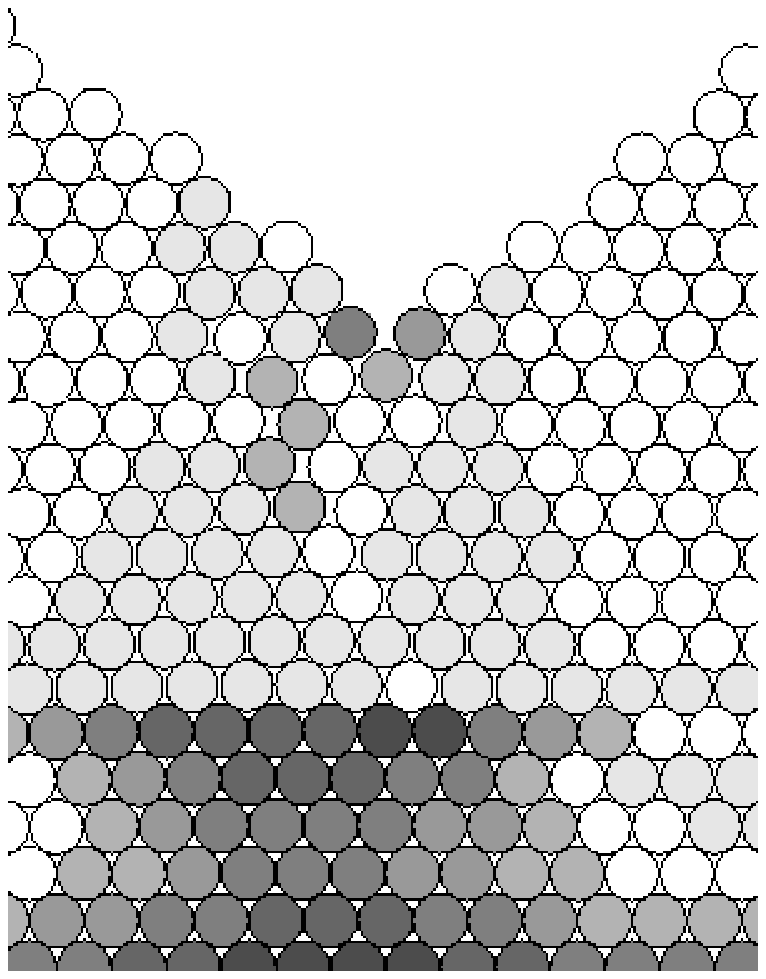}
     \makebox[4mm]{}
     \includegraphics[scale=0.30]{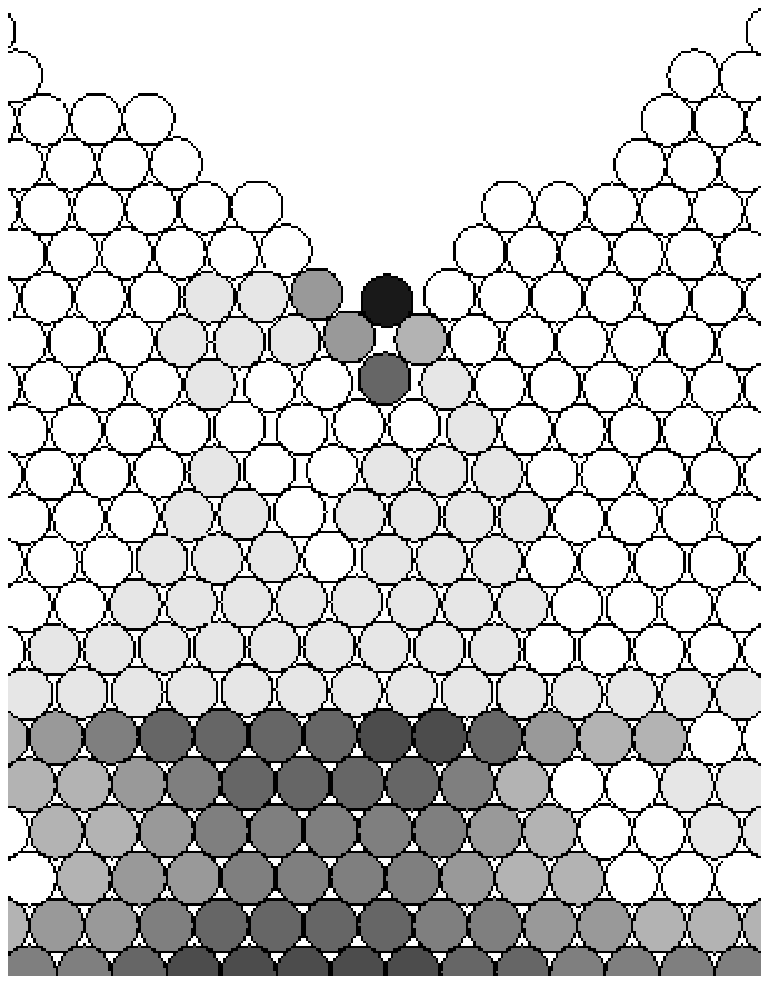}
     \makebox[4mm]{}
     \includegraphics[scale=0.30]{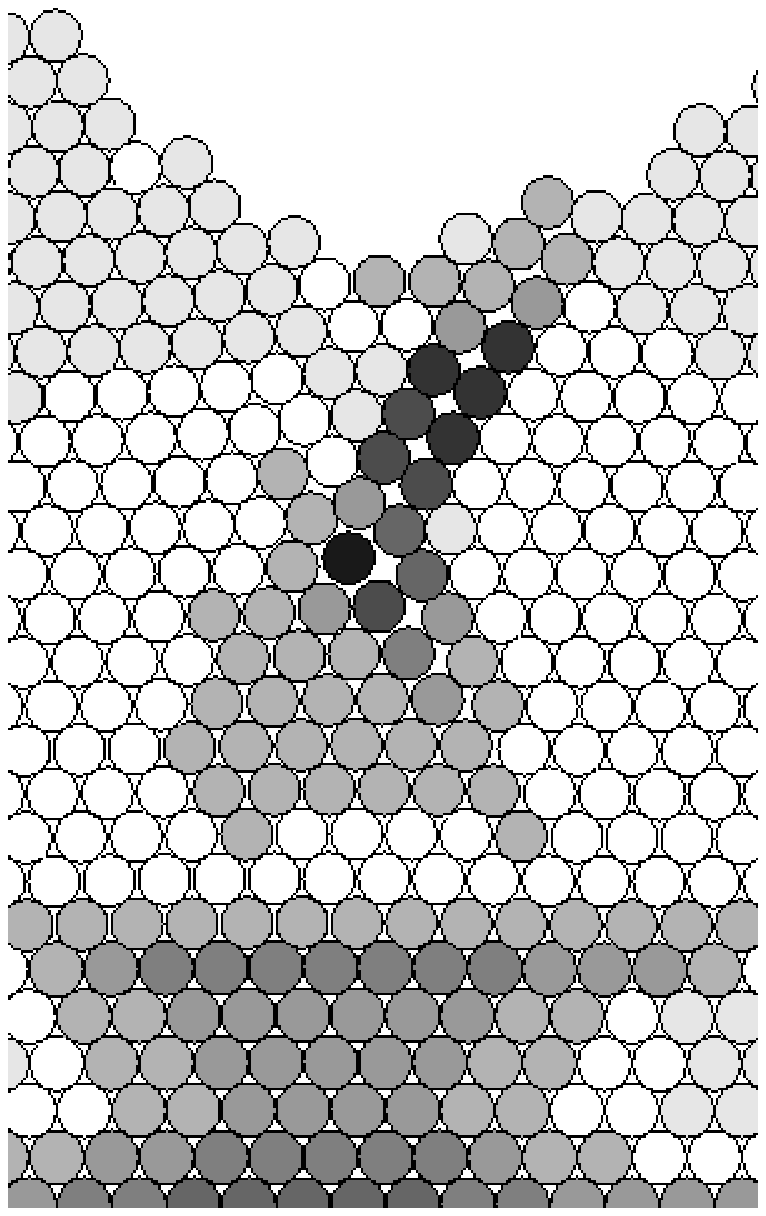}
     \makebox[1mm]{}
     }
     \caption{ \label{dislforms}
     The emergence of a  misfit dislocation
     in the model without downhill funneling \cite{epldislo}
     for $\epsilon=-5\%$, other details as in Fig.\ \ref{manydisl}. The snapshots
     correspond to a mean film thickness of $12 ML, 13 ML$ and $18 ML$, 
     respectively.
    }    
    \end{center}
    \end{figure}

For smaller misfits, one observes a more interesting
evolution of the surface: initially, a pseudomorphic 
adsorbate film grows with the same mean lateral atomic distances 
$\bar{a}_{lat} \propto \sigma_s$ as in
the substrate.  At a rather well defined film thickness,
misfit dislocations appear everywhere on the surface, which
are later overgrown. Snapshots of the surface evolution
in the vicinity of a single dislocation are shown in Figure 
\ref{dislforms} for $\epsilon=-5\%$.

In our most recent studies of the low misfit regime, 
we have slightly amended the model 
in comparison with Section \ref{model}.
Deposited particles perform a so-called 
\textit{downhill funneling\/} upon arrival at the surface: It is
assumed that its kinetic energy enables an arriving atom to move 
to the lowest position in the vicinity of the deposition site.
Similar incorporation effects, which are clearly not thermally
activated, are discussed in \cite{biehl} (this volume) and in 
greater detail in \cite{villain,krug}. 
The funneling process smoothens the surface by reducing the
effect of the strong Schwoebel barrier in the system.
It hence allows for the simulation of thicker films without
a pronounced mound structure as, for instance, in Fig.\ \ref{manydisl}.
Note, however, that  our findings are to a large extent
robust with respect to such modifications.

\begin{figure}[t]
    \begin{center} 
     \includegraphics[scale=0.32]{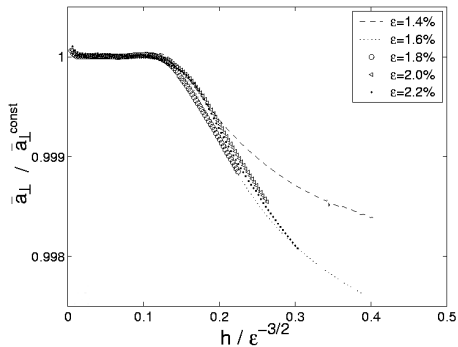}
     \caption{\label{aperp}
     Dislocations: development of the vertical layer spacing in the model with
     {\sl downhill funneling}, see text, with increasing
     film height for various values of the misfit.
     Results were obtained on average over 10
     independent simulation runs and the spacing $\bar{a}_\perp$ is
     rescaled by its initial value. The scaling of the film thickness
     is according to Eq.\ (\ref{powerlaw}) and results in a relatively
     good collapse of the curves close to the critical thickness.
     }
    \end{center}
    \end{figure}

For positive misfits
the adsorbate is laterally compressed and it reacts by assuming 
a vertical lattice spacing  $\bar{a}_\perp$ which is larger than
in its undisturbed bulk.  A simple consideration yields for
small $\epsilon$ in the LJ-system a vertical spacing 
$\bar{a}_\perp^{const} \approx \bar{a}_{lat} \sqrt{3/4} (1 + 4/3 \epsilon)$
of the atomic layers in the pseudomorphic film. 
After misfit dislocations have emerged,
the adsorbate approaches its relaxed undisturbed bulk structure, 
as indicated by the decrease
of  $\bar{a}_\perp$ with the film thickness.
Figure \ref{aperp} shows the result of our simulations
on average over 10 independent runs for
system sizes of at least $L=400$ 
and misfits in the range $1.4\% \leq \epsilon \leq 2.2 \%$.  
The qualitative evolution of $\bar{a}_\perp$ agrees  with 
novel measurements of this quantity in  experimental studies
of II-VI-semiconductor heteroepitaxy \cite{bader}. In a forthcoming
publication the comparison with experimental data will be
presented in greater detail.

Perhaps the most interesting result in this context
concerns the \textit{critical} film thickness $h_c$ at
which the dislocations appear in the system.  
In Fig. \ref{aperp}, it is marked
by the  deviation of $\bar{a}_\perp$ from its
initial value. The rescaling of the thickness with $\epsilon^{-3/2}$
in Fig.\ \ref{aperp} shows a relatively good collapse of the curves
in the relevant phase of growth. 
It is by no means our intention to suggest that, here,
a true {\sl dynamical scaling law} exists as in, e.g., 
kinetic roughening \cite{villain,krug}.
The small range of considered misfits 
would certainly not allow for such a claim.
Nevertheless, our data is consistent with  a critical thickness 
of the form
\begin{equation} \label{powerlaw}
h_c \, \propto \, \epsilon^{-3/2}.
\end{equation}
In the previous study of a slightly modified model 
we considered a different measure of the critical film thickness \cite{epldislo}.
There, we found the same power law behavior for positive and negative
misfits in a much wider range of misfits. Preliminary studies 
yield the same qualitative result for the model with 
other pair potentials in place of the Lennard-Jones interactions.
Hence we believe that the observed dependence is robust with respect
to details of the model and displays a certain degree of universality.

Note that the power law behavior (\ref{powerlaw}) disagrees with the 
results of energy or force balance considerations, as for instance the well-known
and widely accepted relation derived by Matthews and Blakeslee \cite{blakeslee}. 
One can argue, however, that the dislocation formation as observed here 
is a kinetic phenomenon far from equilibrium. 
Indeed, alternative approaches as suggested by Cohen-Solal and co-workers
 predict the power law dependence  (\ref{powerlaw}) for the critical thickness
\cite{cohen1,cohen2}.  Furthermore, experimental data for several IV-IV, III-V, and
II-VI semiconductor systems shows very good agreement with the hypothesized
power law \cite{cohen1,cohen2,pinardi}. 

In forthcoming investigations we will study the dependence 
of the surface evolution on the temperature and deposition rate. 
This should provide further evidence for the
robustness of relation (\ref{powerlaw}) and for the interpretation 
of misfit induced dislocation formation as an activated process. 

\section{Stranski-Krastanov like growth}  \label{stranski}

Obviously, dislocations should dominate the strain relaxation
in sufficiently thick films and for large misfits. 
In material systems 
with relatively small mismatch an alternative effect governs the 
initial growth of very thin films: Instead of growing layer by layer, the
adsorbate aggregates in three-dimensional structures, allowing for
partial relaxation.  The term {\sl 3D-islands\/}
is used to indicate that these structures are spatially separated.
The effect is to be distinguished
from the emergence of {\sl mounds\/} 
due to the Ehrlich-Schwoebel instability \cite{villain,krug}
which also occurs in homoepitaxy.

 At least two distinct growth scenarios display 3D-island for\-mation:
 In {\sl Volmer-Weber\/} growth, the  structures emerge
 immediately upon the substrate before the adsorbate
 even forms a closed layer \cite{villain}.
 The situation resembles the formation of non-wetting droplets of 
 liquid on a surface. It is often observed in systems where 
 adsorbate and substrate are fundamentally different, an  example being
 Pb on a graphite substrate \cite{villain}.

 Here, we will focus on on the so--called {\sl Stranski-Krastanov\/}  (SK)
 growth mode, where 3D-islands are found upon a persistent
 pseudomorphic wetting-layer (WL) of adsorbate material \cite{villain,QD}.
 Most prominent examples for SK--systems are Ge/Si and InAs/GaAs where, as 
 in almost all cases discussed in the literature, the adsorbate
 is under compression in the WL.  

 In order to avoid conflicts with more detailed or more restricted
 definitions of SK growth in the literature, we 
 use the term {\sl  SK-like growth \/} here. 
 It summarizes the following sequence of events
 during the deposition of a few monolayers $(ML)$ of material: 
  \begin{enumerate}          
 \item Initial layer by layer growth of a pseudomorphic, compressed
 adsorbate WL.
 \item The sudden appearance of 3D-islands, marking the so-called 
 2D-3D- or SK-transition at a {\sl kinetic WL thickness\/}  $h_{\rm WL}^{*}$.
 \item Further growth of the 3D-islands, which is
 fed by additional deposition and by incorporation
 of surrounding WL atoms.
 \item The observation of separated 3D-islands of similar shapes and sizes, 
 on top of a WL with reduced {\sl stationary thickness\/} $h_{\rm WL} < h_{\rm WL}^{*}$. 
 \end{enumerate} 

 In order to avoid confusion, and since one might interpret our (1+1)-dim. model
 as a cross section of the full (2+1)-dim. picture,
 we will use the term 2D-3D-transition as usual throughout the following. 

In SK growth a number of effects might play important roles,
including the mixing interdiffusion of materials
or the segregation of compound adsorbates.
These effects are certainly highly relevant in many cases, see several
contributions in \cite{kotrla} and \cite{QD}.
However, SK--like growth is observed in a variety of material systems
which may or may not display these specific features. 
For instance, intermixing or segregation should be irrelevant in the somewhat
{\sl exotic\/} case of large organic molecules like PTCDA 
deposited on a metal substrate.
Nevertheless, this system shows SK--like growth
according to the above  definition \cite{moritz}. 

This  very diversity of SK--systems gives rise to the 
hope that this growth scenario might be explained
in terms of a few basic mechanisms.
Accordingly, it should be possible to capture
and identify these universal features in relatively simple model systems
without aiming at the reproduction of material specific details.  
This hope motivated the investigation of SK-growth in the frame
of our off-lattice model. 

An important modification beyond the description in Section \ref{model}
concerns the interlayer diffusion of adatoms.
As argued above, the Ehrlich-Schwoebel effect
would be much less pronounced in the physical $(2+1)$-dim.\ situation.
In our investigation of the SK-like scenario we  remove the ES-barrier
for all interlayer diffusion events at terrace edges {\sl by hand\/}.  One 
motivation is the above mentioned over-estimation. 
More importantly, we wish to investigate strain induced 
island formation without interference of the ES instability.
Note that the latter leads to the formation of mounds even in homoepitaxy 
\cite{villain,krug}.

In order to favor the emergence of a wetting layer, we
set $U_{s} > U_{as} > U_{a}$ in our model. 
As an example we have used $U_{s}=1.0eV$, $U_{a}=0.74eV$, and
$U_{as} = 0.86 eV$ accordingly.
If not otherwise specified, 
we consider a positive misfit of $\epsilon=4\%$ in the following.
We have demonstrated before \cite{epldislo}, cf. Section \ref{dislocations},
that strain relaxation through dislocations  
is not expected for  $\epsilon = 4\%$ within the first few adsorbate layers.  
Indeed, no misfit dislocations were 
observed in the simulations presented here. 

In the simulation we realize a situation very similar to
many experiments: 
a fixed amount of adsorbate material, corresponding to
$4ML$, is deposited at a constant rate $R_d$.  After
deposition ends,  we allow for a short relaxation period,
in which atoms can still diffuse on the surface. Note
however, that the essential surface properties are already
determined during growth. 
    \begin{center}
    \begin{figure} 
     \includegraphics[scale=0.7]{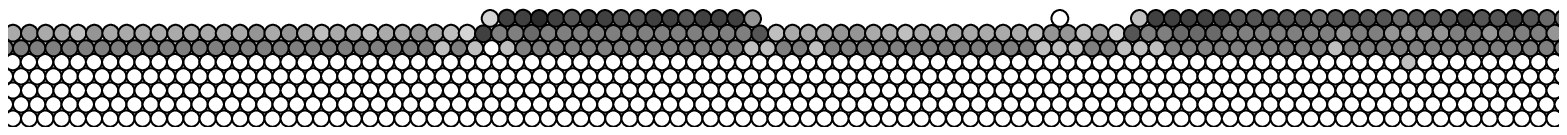}   
     \includegraphics[scale=0.7]{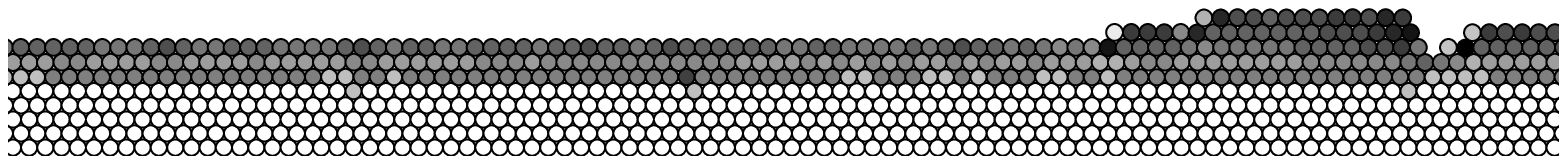}  
     \includegraphics[scale=0.7]{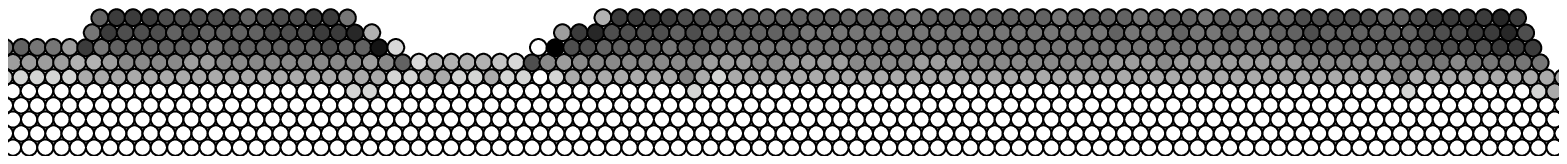} 
     \includegraphics[scale=0.7]{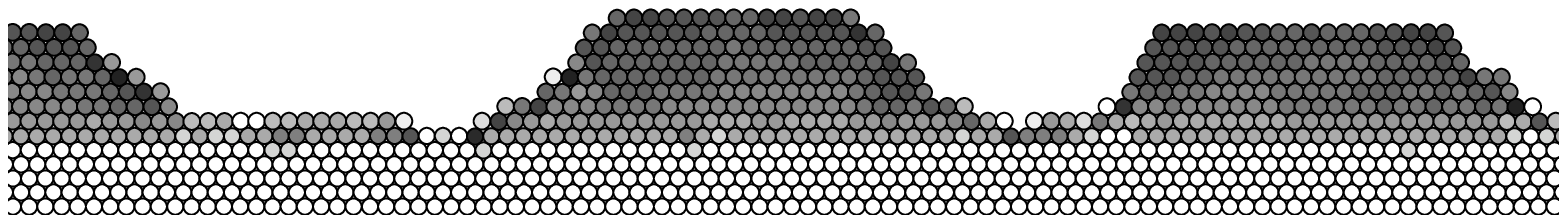}
    \caption{\label{skfilm}
         SK-like growth: A section of a simulated crystal with six layers
         of substrate as obtained for  $\epsilon=+4\%$ and
         $R_d = 7.0 ML/s$  at $T=500K$ after deposition
         of (top to bottom) $1.5 ML, 2.3 ML, 3.0 ML$ and
         $4.0 ML$.  In the second snapshot  the
         critical WL thickness $h_{WL}^* \approx 2.3 ML$
         is reached and island formation sets in.  Note how
         smaller islands grow and larger structures break up.
         Note also that WL particles contribute to the
         growth of islands. Eventually,
         the well separated structures are located 
         on a stationary WL with, here, $h_{WL} \approx 1$.
         The darker a particle is displayed, the larger is the 
         average distance from its nearest neighbors. White represents 
         $\sigma_s$ whereas $\sigma_a$ corresponds to a dark grey
         on this scale.} 
    \end{figure}
    \end{center}

Our model displays a behavior along the lines of 
our operative definition of SK-like growth. The snapshots of
an example simulation run in Figure \ref{skfilm} correspond
precisely to the four stages outlined above, 
illustrating mpeg-movies of sample simulations are available at 
\cite{homepage}.

Several properties of surface diffusion in our model
are  discussed in \cite{nato} in greater detail. 
Here, we shortly summarize the essential features:
\begin{itemize}
\item[a)] Adsorbate diffusion right on the substrate
      is relatively slow, due to the fact that  
      $U_{as}$ is quite large and satisfies $U_{as} > U_{a}$. 
\item[b)] Adsorbate diffusion on a strained WL
      is relatively fast. The diffusion barrier
      decreases with the thickness, but essentially
      saturates at 3 or $4ML$ in our model, where
      the interaction with the substrate becomes
      negligible. 
\item[c)] An adsorbate particle on top of an
          existing island is subject to a strong
          diffusion bias towards the island center.
          This result was already reported in \cite{schroederwolf}.
          The bias is due to the partial relaxation in
          the island and unrelated to the Ehrlich-Schwoebel effect,
          which has been eliminated in our model. 
\end{itemize}

The first two effects, (a) and (b), 
clearly favor and stabilize the existence 
of a wetting layer. Qualitatively the same  relation 
is obtained in experimental investigations of the Ge/Si system
\cite{voigtlaender} and \cite{QD}.
With the model parameters specified above
we find an activation barrier of  approximately $0.57 eV $
for directly on the substrate 
and roughly $0.47 eV$ for diffusion on the first wetting layer.

On the contrary, the diffusion bias (c) stabilizes existing
islands by essentially confining adatoms to their top terraces. 
Of course this is the case for particles deposited onto the
island. More importantly, we find a significant contribution
of particles that jump upward, from the WL atop an island. 
The rate for such processes is quite small, generally,
but becomes significant close to the SK-transition, see 
\cite{nato} for details. 

In simulations with different deposition rates 
we observe that the kinetic WL thickness increases
with $R_d$ \cite{eplstranski,nato}.
If  the formation of second or third layer nuclei
by freshly deposited particles was the driving force, one would expect
more frequent nucleation and an earlier 2D-3D-transition at higher
growth rates. 
We conclude that, in the main, upward jumps trigger the SK-transition. 
Further evidence and a more detailed discussion of this important point
is given in \cite{nato}. 

Our investigations suggest the 
following picture of the Stranski-Krastanov transition: 
The diffusion properties, (a) and (b), favor the formation of
a wetting layer and stabilize it. 
As the film grows, strain accumulates in the film and 
upward hops from the WL become more probable. The precise spatial
modulation of $R_d$ in the film  presumably determines the final
island sizes and will  be studied  in forthcoming investigations. 
Once multilayer islands have emerged they are stabilized by the diffusion 
bias (c). 
 
After the 2D-3D-transition,  islands grow by incorporating newly deposited 
material, but also by consumption of the surrounding WL. 
Note that very large islands are observed to split by means of 
upward diffusion events onto their top layer, cf. Fig.\ \ref{skfilm}.
The migration of WL particles towards and onto the islands can well 
extend into the relaxation period after deposition ends. 
Eventually, a stationary WL thickness $h_{\rm WL} \approx 1$ is observed
in our example scenario with $U_{as} \approx 0.86 eV$.  By increasing the 
strength of the adsorbate/substrate interaction  we can achieve, e.g.,
$h_{\rm WL} \approx 2$ for $U_{as} \approx 2.7 eV$,
but it is difficult to stabilize a greater stationary
thickness. This effect is related to the effective 
short range nature of the LJ-potential, which is 
very weak for distances larger than $3 \sigma$. 
With, e.g., long-range or  multi-particle interactions,
the model should yield greater WL thicknesses. 

\begin{figure} 
    \begin{center}
     \includegraphics[scale=0.36]{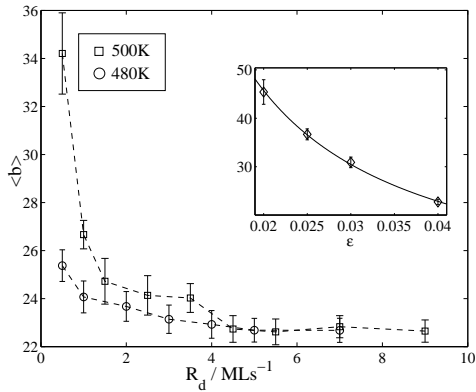}
    \caption{ \label{saturate}
     Average base size $\left\langle b \right\rangle$   
     of multilayer islands as a function of $R_d$ at $T=480$ and $T=500K$, together
     with standard error bars. Results were obtained on average over
     15 independent simulation runs.  The inset shows the result for $T= 500K,R_d= 4.5 ML/s$  
     and different misfit parameters, the solid line corresponds to 
     $\left\langle b \right\rangle = 0.91/\epsilon$.  } 
    \end{center}
    \end{figure}

Finally, we discuss some properties of the emerging islands or SK-Dots.
Figure \ref{saturate} displays the average lateral size,
measured as the number of particles in the island bottom layer.  
The results shown here were obtained at the end of a short relaxation
period with $R_d=0$. Whereas the mean values do not change significantly,
fluctuations are observed to decrease in this phase.

We observe for two different temperatures that the island size decreases with
increasing deposition rate. This is in accordance with several experimental
observations \cite{seifert}. However,  the size becomes constant and
independent of $T$ for large enough deposition flux. A corresponding
behavior is found for the island density and their lateral spacing,
which hints at a considerable degree of spatial ordering \cite{eplstranski,nato}.
This saturation behavior further demonstrates the relevance of upward hops.
Standard arguments \cite{villain,krug} show that a dominant  
aggregation of deposited particles on top terraces would yield an
island density that continues to increase with $R_d$.

The inset in Fig.\ \ref{saturate} displays the mean island size
for $T=500K$ and $R_d= 4.5 ML/s$, i.e. in the  saturation regime.
Our result is consistent with a simple power law of the form $\left\langle b \right\rangle
 \propto 1/\epsilon$. Very far from equilibrium, the only relevant length scale
 in the system appears to be, again, the relative periodicity $1/\epsilon$ of 
adsorbate and substrate lattice.

\section{Summary and outlook}  \label{outlook}

Despite its conceptual simplicity and the small number of
parameters, our model reproduces several features of heteroepitaxial growth.
Strain effects are incorporated in a natural fashion as they emerge
directly from the interaction of particles.
For instance, we have demonstrated that misfit dislocations appear in the system when
the film thickness exceeds a characteristic value. This characteristic
height displays a power law behavior on the lattice misfit. 

We furthermore believe
that, with appropriately chosen interaction parameters,  our model is
capable of reproducing the three essential growth modes: 
extended layer by layer growth 
(for very small misfits), Volmer-Weber for $U_{as} < U_a$ 
and, as demonstrated already, SK-growth for $U_{as}>U_a$.  
The small number of parameters  should allow for determining 
the corresponding {\sl phase diagram\/} of growth modes in the frame of
our model system.

Besides the exploration of the available parameter space, we intend to 
extend the model conceptually in several directions. 
As just one example,
intermixing and seggregation should be included to study the relevance
of these  effects in the Quantum Dot formation.   

In order to test the potential universality of our findings, we
will introduce different types of interactions in our model. 
Preliminary results for,e.g., Morse potentials  \cite{chemistry}
and modifications of  LJ-potentials confirm our results in the
context of dislocation formation. 
Ultimately, we will extend our model to the physical
case of growth in $(2+1)$ dimensions and to realistic
empirical potentials for metals or semi-conductors. 
As a first step, we are currently investigating the sub-monolayer
regime in strained heteroepitaxy of fcc materials.   

In a sense,  the off-lattice KMC method provides a link between traditional,
lattice based methods  and Molecular Dynamics. Hence, it might play
a significant role in the further development  of the multiscale approach.


\subsection*{Acknowledgment}
M.B.\  would like to thank the organizers and all participants of the 
MFO Mini-Workshop 
 on {\sl Multiscale Modelling in Epitaxial Growth\/}
for the most stimulating atmosphere and many
useful discussions. One of us (F.M.) was supported by the Deutsche
Forschungsgemeinschaft. 
\end{document}